\documentstyle[11pt,aaspp4]{article}

\def\simg{\mathrel{\hbox{\rlap{\lower.55ex \hbox {$\sim$}}
                   \kern-.3em \raise.4ex \hbox{$>$}}}}
\def\siml{\mathrel{\hbox{\rlap{\lower.55ex \hbox {$\sim$}}
                   \kern-.3em \raise.4ex \hbox{$<$}}}}
\def\Mesz{M\'esz\'aros~}
\def\Pacz{Paczy\'nski~}
\def\Kluz{Klu\'zniak~}
\def\beq{\begin{equation}}
\def\enq{\end{equation}}
\def\bea{\begin{eqnarray}}
\def\ena{\end{eqnarray}}
\def\bec{\begin{center}}
\def\enc{\end{center}}
\def\E52{E_{52}}
\def\r13{r_{13}}
\def\et2{\eta_2}

\def\th2{\theta^2}
\def\t5{t_5}
\def\kap3{\kappa_3}

\def\calE{{\cal E}}
\def\Omj{\Omega_j}
\def\msun{M_\odot}
\def\Mbh{M_{bh}}
\def\eps{\epsilon}

\begin{document}

\title{ Energetics and Beaming of Gamma Ray Burst Triggers~\footnote{Manuscript prepared for New Astronomy, David N. Schramm memorial volume}  }

\author{ \Mesz, P.$^1$, Rees, M.J.$^2$ \& Wijers, R.A.M.J.$^{2,3}$ }

\noindent
$^1$Dpt. of Astronomy \& Astrophysics, Pennsylvania State University,
University Park, PA 16803 \\
$^2$Institute of Astronomy, University of Cambridge, Madingley Road, Cambridge
CB3 0HA, U.K.\\
$^3$Dpt. of Physics \& Astronomy, SUNY, Stony Brook, NY 11794-3800\\
\bec Date :~~{ 9/03/98} \enc

\begin{abstract}
A wide range of mechanisms have been proposed to supply the energy
for gamma-ray bursts (GRB) at cosmological distances. It is a common
misconception that some of these, notably NS-NS mergers, cannot meet
the energy requirements suggested by recent observations.  We show here
that GRB energies, even at the most distant redshifts detected, are
compatible with current binary merger or collapse scenarios involving
compact objects. This is especially so if, as expected, there is a
moderate amount of beaming, since current observations constrain the
energy per solid angle much more strongly and directly than the
total energy.  All plausible progenitors, ranging from NS-NS mergers to
various hypernova-like scenarios, eventually lead to the formation of a
black hole with a debris torus around it, so that the extractable energy
is of the same order, $10^{54}$ ergs, in all cases.  MHD conversion of
gravitational into kinetic and radiation energy can significantly increase
the probability of observing large photon fluxes, although significant
collimation may achieve the same effect with neutrino annihilation in
short bursts.  The lifetime of the debris torus is dictated by
a variety of physical processes, such as viscous accretion and various 
instabilities; these mechanisms dominate at different stages in the 
evolution of the torus and provide for a range of gamma-ray burst lifetimes.

\end{abstract}


\vfill\eject

\section{Introduction}

The discovery of afterglows in the last year has moved the investigation of
gamma-ray bursts (GRB) to a new plane. It not only has opened the field to new 
wavelengths and extended observations to longer time scales, making the 
identification of counterparts possible, but also provided confirmation for 
much of the earlier work on the fireball shock model of GRB, in which the 
$\gamma$-ray emission arises at radii of $10^{13}-10^{15}$ cm (Rees \& \Mesz 
1992, 1994, \Mesz \& Rees 1993, \Pacz \& Xu 1994, Katz 1994, Sari \& Piran 
1995). In particular, 
this model led to the prediction of the quantitative nature of the signatures 
of afterglows, in substantial agreement with subsequent observations (\Mesz 
\& Rees 1997a, Costa et~al.\ 1997, Vietri 1997a, Tavani 1997, Waxman 1997; 
Reichart 1997, Wijers et~al.\ 1997).
More recently, significant interest was aroused by the report of an afterglow
for the burst GRB971214 at a redshift $z=3.4$, whose fluence corresponds to 
a $\gamma$-ray energy of $10^{53.5} (\Omega_\gamma /4\pi)$ erg (Kulkarni 
et~al.\ 1998). There is also possible evidence that some fraction of the
detected afterglows may arise in relatively dense gaseous environments. This 
is suggested, e.g. by evidence for dust in GRB970508 
(Reichart 1998), the absence of an optical afterglow and presence of
strong soft X-ray 
absorption in GRB 970828 (Groot et~al.\ 1997, Murakami et~al.\ 1997), the lack 
an an optical afterglow in the (radio-detected) afterglow of GRB980329 (Taylor 
et~al.\ 1998), etc.  This has led to the suggestion that ``hypernova" models 
(\Pacz\ 1998, Fryer \& Woosley 1998) may be responsible, since hypernovae are 
thought to involve the collapse of a massive star or its merger with a compact 
companion, both of which would occur on time scale short enough to imply a 
burst within the star forming region. By contrast, neutron star - neutron star
(NS-NS) or neutron star - black hole (NS-BH) mergers 
would lead to a similar BH plus debris torus system and roughly the same total 
energies (a point not generally appreciated), but the mean distance traveled 
from birth is of order several kpc (Bloom, Sigurdsson \& Pols 1998), leading 
to a burst presumably in a less dense environment. The fits of Wijers \& 
Galama (1998) to the observational data on GRB 970508 and GRB 971214 in fact
suggest external densities in the range of 0.04--0.4 cm$^{-1}$, which would be 
more typical of a tenuous interstellar medium. In any case, 
while it is at present unclear which, if any, of these progenitors is 
responsible for the bulk of GRB, or whether perhaps different progenitors 
represent different subclasses of GRB, there is general agreement that they 
all would be expected to lead to the generic fireball shock scenario 
mentioned above.

\section{ Trigger Mechanisms and Black Hole/Debris Torus Systems}

The first detailed investigations of the disruption of a NS in a merger with 
another NS or a BH were carried out by Lattimer \& Schramm (1976), and the 
significance of this work for GRB has only recently started to be appreciated.
It has become increasingly apparent in the last few years  that {\it all}
plausible GRB progenitors suggested so far (e.g. NS-NS or NS-BH mergers, 
Helium core - black hole [He/BH] or white dwarf - black hole [WD-BH] mergers, 
and a wide category labeled as hypernova or collapsars including failed 
supernova Ib [SNe Ib], single or binary Wolf-Rayet [WR] collapse, etc.) are 
expected to lead to a BH plus debris torus system. An important point is that 
the overall energetics from these various progenitors do not differ 
by more than about one order of magnitude. 

Two large reservoirs of energy are available in principle: the binding energy 
of the orbiting debris, and the spin energy of the black hole (\Mesz \& Rees, 
1997b). The first can provide up to 42\% of the rest mass energy of the disk, 
for a maximally rotating black hole, while the second can provide up to 29\% 
of the rest mass of the black hole itself.
The $\nu\bar\nu \to e^+ e^-$ process (Eichler et~al.\ 1989) can tap the thermal 
energy of the torus produced by viscous dissipation. For this mechanism to
be efficient, the neutrinos must escape before being advected into the hole;
on the other hand, the efficiency of conversion into pairs (which scales with
the square of the neutrino density) is low if the neutrino production is too
gradual. Typical estimates suggest a  fireball of $\siml 10^{51}$ erg 
(Ruffert et al 1997, Popham, Woosley \& Fryer 1998), except perhaps in the 
``collapsar" or failed SN Ib case where Popham et~al.\ (1998) estimate 
$10^{52.3}$ ergs for optimum parameters. If the fireball is collimated
into a solid angle $\Omj$ then of course the apparent ``isotropized" energy 
would be larger by a factor $(4\pi/\Omj)$ , but unless $\Omj$ is $\siml 10^{-2}
-10^{-3}$ this may fail to satisfy the apparent isotropized energy of 
$10^{53.5}$ ergs implied by a redshift $z=3.4$ for GRB 971214.
An alternative way to tap the torus energy is through dissipation of magnetic 
fields generated by the differential rotation in the torus (\Pacz\ 1991, 
Narayan, \Pacz \& Piran 1992, \Mesz \& Rees 1997b, Katz 1997). Even 
before the BH forms, a NS-NS merging system might lead to winding up of the 
fields and dissipation in the last stages before the merger (\Mesz \& Rees 
1992, Vietri 1997a). 
The above mechanisms tap the energy available in the debris torus or disk.
However, a hole formed from a coalescing compact binary is guaranteed to be 
rapidly spinning, and, being more massive, could contain more energy than
the torus; the energy extractable in principle through MHD coupling to the 
rotation of the hole by the Blandford \& Znajek (1977) effect could then be 
even larger than that contained in the orbiting debris (\Mesz \& Rees 1997b, 
\Pacz\ 1998). Collectively, any such MHD outflows have been referred to as 
Poynting jets.

The various progenitors differ only slightly in the mass of the BH and
that of the debris torus they produce, and they may differ more markedly
in the amount of rotational energy contained in the BH. Strong magnetic
fields, of order $10^{15}$ G, are needed needed to carry away the
rotational or gravitational energy in a time scale of tens of seconds
(Usov 1994, Thompson 1994).  If the magnetic fields do not thread the BH,
then a Poynting outflow can at most carry the gravitational binding energy
of the torus. For a maximally rotating and for a non-rotating BH this is
0.42 and 0.06 of the torus rest mass, respectively. The torus or disk mass
in a NS-NS merger is $M_d\sim 0.1\msun$ (Ruffert \& Janka 1998), and for a
NS-BH, a He-BH, WD-BH merger or a binary WR collapse it may be estimated at  
$M_d \sim 1\msun$ (\Pacz\ 1998, Fryer \& Woosley 1998).  In the HeWD-BH merger
and WR collapse the mass of the disk is uncertain due to lack of
calculations on continued accretion from the envelope, so $1\msun$ is just
a rough estimate. The largest energy reservoir is therefore, `prima
facie', associated with NS-BH, HeWD-BH or binary WR collapse, which have
larger disks and fast rotation, the maximum energy being $\sim 8 \times
10^{53} \eps (M_d/\msun)$ ergs; for the failed SNe Ib (which is a slow
rotator) it is $\sim 1.2\times 10^{53}\eps (M_d/\msun)$ ergs, and for the
(fast rotating) NS-NS merger it is $\sim 0.8\times 10^{53} \eps (M_d/0.1
\msun) $ ergs, where $\eps$ is the efficiency in converting gravitational
into MHD jet energy.  Conditions for the efficient escape of a
high-$\Gamma$ jet may, however, be less propitious if the ``engine" is
surrounded by an extensive envelope.

If the magnetic fields in the torus thread the BH, the rotational energy
of the BH can be extracted via the B-Z (Blandford \& Znajek 1977)
mechanism (\Mesz \& Rees 1997b). The extractable energy is
$\eps f(a)\Mbh c^2$,
where $\eps$ is the MHD efficiency factor and $a = Jc/G M^2$ is the
rotation parameter, which equals 1 for a maximally rotating black hole.
$f(a)=1-\sqrt{\frac{1}{2}[1+\sqrt{1-a^2}]}$ is small unless $a$
is close to 1, where it sharply rises to its maximum value $f(1)=0.29$,
so the main requirement is a rapidly rotating black hole, $a \simg 0.5$.
For a maximally rotating BH, the extractable energy is therefore
$0.29 \eps \Mbh c^2 \sim 5\times 10^{53} \eps (\Mbh/\msun)$ ergs.
Rapid rotation is essentially guaranteed in a  NS-NS merger, since
the radius (especially for a soft equation of state) is close to that of a
black hole and the final orbital spin period is close to the required
maximal spin rotation period. Since the central BH will have a mass of
about $2.5 \msun$ (Ruffert \& Janka 1998), the NS-NS system can thus power 
a jet of up to $\sim 1.3 \times 10^{54} \eps (\Mbh/2.5\msun)$ ergs.  The
scenarios less likely to produce a fast rotating BH are the NS-BH merger
(where the rotation parameter could be limited to $a \leq M_{ns}/\Mbh$,
unless the BH is already fast-rotating) and the failed SNe Ib (where the
last material to fall in would have maximum angular momentum, but the
material that was initially close to the hole has less angular momentum).
A maximal rotation rate may also be possible in a He-BH merger, depending
on what fraction of the He core gets accreted along the rotation axis as
opposed to along the equator (Fryer \& Woosley 1998), and the same should
apply to the binary fast-rotating WR scenario, which probably does not
differ much in its final details from the He-BH merger. For a fast
rotating BH of $3\msun$ threaded by the magnetic field, the maximal energy
carried out by the jet is then $\sim 1.6\times 10^{54} \eps
(\Mbh/3\msun)$ ergs.

Thus in the accretion powered jet case the total energetics between the
various models differs at most by a factor 20, whereas in the rotationally
(B-Z) powered cases they differ by at most a factor of a few, depending on the
rotation parameter. For instance, even allowing for low total efficiency 
(say 30\%), a NS-NS merger whose jet is powered by the torus binding energy 
would only require a modest beaming of the $\gamma$-rays by a factor 
$(4\pi/\Omj)\sim 20$, or no beaming if the jet is powered by the B-Z mechanism, 
to produce the equivalent of an isotropic energy of $10^{53.5}$ ergs.  The 
beaming requirements of BH-NS and some of the other progenitor scenarios are 
even less constraining.

\section{ Intrinsic Time scales } 

A question which has remained largely unanswered so far is what determines  the
characteristic duration of bursts, which can extend to tens, or even hundreds, 
of seconds.  This is of course very long in comparison with the dynamical or 
orbital time scale for the ``triggers" described in section 2. While bursts 
lasting hundreds of seconds can easily be derived from a very short, impulsive
energy input, this is generally unable to account for a large fraction
of bursts which show complicated light curves. This hints at the desirability 
for a ``central engine" lasting much longer than a typical dynamical time scale. 
Observationally (Kouveliotou et~al.\ 1993) the short ($\siml 2$ s) and
long ($\simg 2$ s) bursts appear to represent two distinct subclasses, and
one early proposal to explain this was that accretion induced collapse
(AIC) of a white dwarf (WD) into a NS plus debris might be a candidate for
the long bursts, while NS-NS mergers could provide the short ones (Katz \&
Canel 1996).  As indicated by Ruffert et~al.\ (1997), $\nu\bar\nu$
annihilation will generally tend to produce short bursts $\siml 1$ s in
NS-NS systems, requiring collimation by $10^{-1}-10^{-2}$, while Popham,
Woosley \& Fryer (1998) argued that in collapsars and WD/He-BH systems
longer $\nu\bar\nu$ bursts may be possible. Longer bursts however imply
lower $e^\pm$ conversion efficiency, so the observed fluxes could then be
explained only if the jets were extremely collimated, by at least
$10^{-3}-10^{-4}$. We outline here several possible mechanisms, within the
context of the basic compact merger or collapse scenario leading to a BH
plus debris torus, which can lead to an adequate energy release on such
time scales.

  If the trigger of a long-duration burst involves a black hole, then an
acceptable model requires that the  surrounding torus should not 
completely drain into the hole, or be otherwise dispersed, on too short
a  time scale. There have been some discussions in the literature of
possible 'runaway instabilities' in relativistic tori (Nishida et~al.\
1996, Abramowicz, Karas \& Lanza 1997, Daigne \& Mochkovitch 1997):
these are analogous to the runaway Roche lobe overflow predicted, under
some conditions, in binary systems. These instabilities can be virulent
in a torus where the specific angular momentum is uniform throughout,
but are inhibited by a spread in angular momentum.  In a torus that was
massive and/or thin enough to be self-gravitating, bar-mode gravitational
instabilities could lead to further  redistribution of angular momentum 
and/or to energy loss by gravitational radiation within only a few orbits.
Whether a torus of given mass is dynamically unstable depends on its 
thickness and stratification, which in turn depends on internal viscous 
dissipation and neutrino cooling. 

   The disruption of a neutron star (or any analogous process) is
almost certain to lead to a situation where violent instabilities
redistribute mass and angular momentum within a few dynamical time scales
(i.e. in much less than a second). A key issue  for gamma ray burst models
is the nature of the surviving debris after these violent processes are
over: what is the maximum mass  of a remnant disc/torus which is immune
to very violent instabilities, and which can therefore in principle
survive for long enough to   power the bursts? 

   \subsection{Magnetic torques and viscosity}

Differential rotation may amplify magnetic fields until magnetic viscosity
dominates neutrino viscosity. Moreover, the torques associated with a
large scale magnetic field  may also extract energy and angular momentum
by driving a relativistic outflow.  If the trigger is to generate the
burst energy, over a period 10--100 sec,  via Poynting flux --- either
through a relativistic wind 'spun off' the torus or via  the
Blandford-Znajek  mechanism --- the required  field is a few times 
$10^{15}$\,G. A weaker field would extract inadequate power; on the other
hand, if the large-scale field were even stronger, then the energy would
be dumped too fast to account for the longer complex bursts.

How plausible are fields of this strength? \Kluz and Ruderman (1998) point
out that, starting with $10^{12}$ G, it only takes of order a second for
simple winding to amplify the field to $10^{15}$ G; they argue further
that magnetic stresses would then be strong enough for flares to break
out. But amplification in a newly-formed torus could well occur more
rapidly, for instance via convective instabilities, as  in a newly formed
neutron star (cf.\ Duncan \& Thompson 1992, Thompson 1994). Such fields
can build up on very short time scales, or order $\sim$ few ms; however,
convective overturning motions should stop after the disk has cooled by
neutrino emission below a few MeV.  The latter is generally estimated to
be of order a few seconds (Ruffert et al, 1997). But azimuthal magnetic
fields can also be generated  via the Balbus-Hawley mechanism.  The
nonlinear evolution and/or reconnection of such fields as they become
buoyant can then lead to poloidal components at least of order $\simg
10^{15}$ G.  Indeed, it is not obvious why the fields cannot become even
higher. Note that the virial limit  is $B_v \sim 10^{17}$ G.

After magnetic fields have built up to some fraction of the equipartition
value with the shear motion, a magnetic viscosity develops. Assuming
that $B_rB_\phi\sim B^2$, it can be characterized in the usual way
by the parameter $\alpha\sim B^2/(4\pi \rho v_s^2 ) \sim 10^{-1} B_{15}^2
\rho_{13}^{-1} T_9^{-1}$. This viscosity continues 
operating also after cooling has led
to the disappearance of neutrino viscosity.  Assuming a value of
$\alpha=0.1$, a BH mass 3 $\msun$ and outer disk radius equal to the Roche
lobe size, Popham et~al.\ (1998) estimate ``viscous" life times of 0.1 s
for NS/BH-NS, 10--20 s for a collapsar (failed SN Ib or rotating WR), and
15--150 s for WD-BH and He-BH systems (although fields of $10^{15}$ G may
be more difficult to support in He-BH systems).

A magnetic field configuration capable of powering the bursts is likely to
have a large scale structure. Flares and instabilities occurring on the
characteristic (millisecond) dynamical time scale would  cause substantial
irregularity or intermittency in the overall outflow that would manifest
itself in internal shocks (Rees \& \Mesz, 1994) There is thus no problem
in principle in accounting for sporadic large-amplitude variability, on
all time scales down to a millisecond, even in the most long-lived
bursts.  Note also that it only takes a residual cold disk of
$10^{-3}\msun$ to confine  a field of $10^{15}$ G, which can extract
energy from the black hole via the Blandford-Znajek mechanism.  Even if
the evolution time scale for the bulk of the debris were no more than a
second, enough may remain to catalyse the extraction of enough energy from
the hole to power a long-lived burst.

      \subsection{Double peaked bursts}

There are at least two  mechanisms which might lead to a delayed
``second" burst (or a double humped burst). One possibility is  that a merger
leads to a central NS, temporarily stabilized by its fast
rotation, with a disrupted debris torus around it, which produces a burst
powered by the accretion energy and the magnetic fields generated by the
shear motions. After the NS has radiated enough of its angular momentum,
and accreted enough matter to overcome its centrifugal support, it collapses
to a BH, leading to a second burst, and second cycle of energy extraction
(either from the disk or from the BH via B-Z). In both cases, the time scale
between bursts should be between a few to few tens of seconds.

The other possibility for a delayed second burst may arise in merging NS
of very unequal masses. As the smaller one fills its Roche lobe and losses
mass, the larger NS (which may also collapse to a BH) is surrounded by the
gas acquired from its companion, producing a burst as above. Eventually
the less massive donor comes under the critical mass for deleptonization,
and this leads to an explosion (e.g. Eichler et~al.\ 1989). Starting from
a configuration with about $0.1\msun$ which losses mass to its companion,
Sumiyoshi et~al.\.  (1998) (see also \Kluz \& Lee 1998, Portegies Zwart
1998) find that the explosion occurs in a time scale of about 20 s.
The importance of this process depends on the poorly known distribution
of NS-NS binary mass ratios, and on whether the mass transfer between
neutron stars of nearly equal mass can be stable.

\section{ Isotropic or Beamed Outflows? }

\noindent
{\it Conversion into relativistic outflow. }
Even if the outflow is not  narrowly beamed,  the energy of a fireball 
would be channeled preferentially along the rotation axis. Moreover, we 
would expect baryon contamination to be lowest near the axis, because angular 
momentum flings material away from the axis, and any gravitationally-bound 
material with low angular momentum falls into the hole.  In hypernova and 
SNe Ib cases without a binary companion, however, the envelope is rotating 
only slowly and thus would not initially have a marked centrifugal funnel;
a funnel might however develop after low angular momentum matter falls into 
the hole along the axis on a free-fall time scale measured from the outer 
radius of the envelope, $t\sim 10^4-10^5$ s. \\

The dynamics are complex. Computer simulations of compact object mergers
and black hole formation can address the fate of the bulk of the matter, but 
there are some key questions that they cannot yet tackle.  In particular, high 
resolution of the outer layers is needed because even a tiny mass fraction of
baryons loading down the outflow severely limits the attainable
Lorentz factor --- for instance a Poynting flux of $10^{52}$ ergs could
not accelerate an outflow to $\Gamma > 100$ if it had to drag more
than $\sim 10^{-4}$ solar masses of baryons with it. Further 2D numerical 
simulations of the merger and collapse scenarios are under way (Fryer \& 
Woosley 1998, Eberl, Ruffert \& Janka 1998, McFayden \& Woosley 1998), 
largely using Newtonian dynamics, and the numerical difficulties are daunting.
There may well be a broad spread of Lorentz factors in the outflow --- close to 
the rotation axis $\Gamma$ may be very high; at larger angles away from
the axis, there may be an increasing degree of entrainment, with a 
corresponding decrease in $\Gamma$.  This picture suggests, indeed, that the 
variety of burst phenomenology could be largely attributable to a standard 
type of event being viewed from different orientations. As discussed in the 
last section, a variety of progenitors can lead to a very similar end result, 
whose energetics are within one order of magnitude from each other. 

\noindent
{\it Basic spherical afterglow model.}
Just as we can interpret supernova remnants even without fully
understanding the initiating explosion, so we may hope to understand
the afterglows of gamma ray bursts, despite the uncertainties
recounted in the previous section.  The simplest hypothesis is that
the afterglow is due to a relativistic expanding blast wave.  The
complex time structure of some bursts suggests that the central
trigger may continue for up to 100 seconds. However, at much later
times all memory of the initial time structure would be lost:
essentially all that matters is how much energy and momentum has been
injected; the injection can be regarded as instantaneous in the
context of the much longer afterglow.

The simplest spherical afterglow model has been remarkably successful at
explaining the gross features of the GRB 970228, GRB970508 and other
afterglows (e.g.\ Wijers et~al.\ 1997). This has led to the temptation to
take the assumed sphericity for granted. For instance, the lack of a break
in the light curve of GRB 970508 prompted Kulkarni et~al.\ (1998a) to
infer that all afterglows are essentially isotropic, leading to their very
large (isotropic) energy estimate of $10^{53.5}$ ergs in GRB 971214. The
multi-wavelength data analysis has in fact advanced to the point where one
can use observed light curves at different times and derive, via
parametric fitting, physical parameters of the burst and environment, such
as the total energy $E$, the magnetic and electron-proton coupling
parameters ${\eps}_B$ and ${\eps}_e$ and the external density $n$ (Waxman
1997, Wijers \& Galama 1998).  However, as emphasized by Wijers \& Galama,
1998, what these fits constrain is only the energy per unit solid angle
$\calE= (E/\Omj)$.

\noindent
{\it Properties of a Jet Outflow.}
An argument for sphericity that has been invoked by observers is that, if
the blast wave energy were channeled into a solid angle $\Omj$ then, as
correctly argued by Rhoads (1997, 1998), one expects a faster decay of
$\Gamma$ after it drops below $\Omj^{-1/2}$. A simple calculation using
the usual scaling laws leads then to a steepening of the flux power law in
time.  The lack of such an observed afterglow downturn in the optical has
been interpreted as further supporting the sphericity of the entire
fireball.  There are several important caveats, however. The first one is
that the above argument assumes a simple, impulsive energy input (lasting
$\siml$ than the observed $\gamma$-ray pulse duration), characterized by a
single energy and bulk Lorentz factor value. Estimates for the time needed
to reach the non-relativistic regime, or $\Gamma < \Omega_j^{-1/2} \siml$
few, could then be under a month (Vietri 1997, Huang, Dai \& Lu 1998),
especially if an initial radiative regime with $\Gamma\propto r^{-3}$
prevails. It is unclear whether, even when electron radiative time scales
are shorter than the expansion time, such a regime applies, as it would
require strong electron-proton coupling (\Mesz, Rees \& Wijers 1998).
Waxman, et~al.\ (1998) have also argued on observational grounds that the
longer lasting $\Gamma \propto r^{-3/2}$ (adiabatic regime) is more
appropriate. Furthermore, even the simplest reasonable departures from a
top-hat approximation (e.g. having more energy emitted with lower Lorentz
factors at later times, which still do not exceed the gamma-ray pulse
duration) would drastically extend the afterglow lifetime in the
relativistic regime, by providing a late ``energy refreshment" to the
blast wave on time scales comparable to the afterglow time scale (Rees \&
\Mesz 1998).  The transition to the $\Gamma < \Omega_j^{-1/2}$ regime
occurring at $\Gamma\sim$ few could then occur as late as six months to
more than a year after the outburst, depending on details of the brief
energy input.  Even in a simple top-hat model, more detailed calculations
show that the transition to the non-relativistic regime is very gradual
($\delta t/t \simg 2$) in the light curve. Also, even though the flux from
the head-on part of the remnant decreases faster, this is more than
compensated by the increased emission measure from sweeping up external
matter over a larger angle, and by the fact that the extra radiation,
which arises at larger angles, arrives later and re-fills the steeper
light curve. The sideways expansion thus actually can slow down the flux
decay (Panaitescu \& \Mesz 1998), rather than making for a faster decay.

As already noted by Katz \& Piran (1997), the ratio $L_\gamma/L_{opt}$ (or 
$L_\gamma / L_x$) can be quite different from burst to burst. The fit of Wijers 
\& Galama for GRB 970508 indicates an afterglow (X-ray energies or softer) 
energy per solid angle ${\cal E}_{52} =3.7$, while at $z=0.835$ with $h_{70}=1$ 
the corresponding $\gamma$-ray ${\cal E}_{52\gamma} =0.63$. On the other hand 
for GRB 971214, at $z=3.4$, the numbers are ${\cal E}_{52} = 0.68$ and 
${\cal E}_{52\gamma}=20$.  The bursts themselves require ejecta with $\Gamma > 
100$. The gamma-rays we receive come only from material whose motion is 
directed within one degree of our line of sight.  They therefore provide no
information about the ejecta in other directions: the outflow could be
isotropic, or concentrated in a cone of angle (say) 20 degrees
(provided that the line of sight lay inside the cone).  At observer
times of more than a week, the blast wave would be decelerated to a
moderate Lorentz factor, irrespective of the initial value. The
beaming and aberration effects are less extreme so we observe
afterglow emission not just from material moving almost directly
towards us, but from a wider range of angles.

The afterglow is thus a probe for the geometry of the ejecta ---
at late stages, if the outflow is beamed, we expect a spherically-symmetric 
assumption to be inadequate; the deviations from the predictions of such a 
model would then tell us about the ejection in directions away from our line 
of sight.  It is quite possible, for instance, that there is relativistic
outflow with lower $\Gamma$ (heavier loading of baryons) in other directions
(e.g.\ Wijers, Rees \& \Mesz 1997); this slower matter could even carry most 
of the energy (\Pacz, 1997). This hypothesis is, if anything, further
reinforced by the fits of Wijers \& Galama (1998) mentioned above.

\noindent
{\it Observational constraints on beaming.}
As discussed above, anisotropy in the burst outflow and emission affects
the light curve at the time when the inverse of the bulk Lorentz factor
equals the opening angle of the outflow. If the critical Lorentz factor
is less than 3 or so (i.e. the opening angle exceeds 20$^\circ$) such
a transition might be masked by the transition from ultrarelativistic
to mildly relativistic flow, so quite generically it would difficult to
limit the late-time afterglow opening angle in this way if it exceeds
20$^\circ$. Since some afterglows are unbroken power laws for over 100
days (e.g.\ GRB\,970228), if the energy input were indeed just a a simple
impulsive top-hat the opening angle of the late-time afterglow at long
wavelengths is probably greater than 1/3, i.e.\ $\Omega_{\rm opt}\simg
0.4$. However, even this still means that the energy estimates from the
afterglow assuming isotropy could be 30 times too high.

The gamma-ray beaming is much harder to constrain directly. The ratio of
$\Omega_\gamma /\Omega_x$ has been considered by Grindlay (1998) using
data from Ariel V and HEAO-A1/A2 surveys, who did not find evidence
for a significant difference between the deduced gamma-ray and X-ray rates,
and concluded that higher sensitivity surveys would be needed to provide 
significant constraints. More promising for the immediate future, 
the ratio $\Omega_\gamma/\Omega_{\rm opt}$ can also be investigated
observationally (see also Rhoads 1997). The rate of GRB with peak
fluxes above 1 ph\,cm$^{-2}$\,s$^{-1}$ as determined by BATSE is about
300/yr, i.e.\ 0.01/sq.~deg/yr. According to Wijers et~al.\.\ (1998) this
flux corresponds to a redshift of 3. If the gamma rays were much more
narrowly beamed than the optical afterglow there should be many `homeless'
afterglows, i.e.\ ones without a GRB preceding them. The transient sky
at faint magnitudes is poorly known, but there are two major efforts
under way to find supernovae down to about $R=23$ (Garnavich et~al.\ 1998,
Perlmutter et~al.\ 1998). These searches have by now covered a few tens of
square degree years of exposure and would be sensitive to afterglows of
the brightness levels thus far observed. It therefore appears that the
afterglow rate is not more than a few times 0.1/sq.~deg/yr. Since the
magnitude limit of these searches allows detection of optical counterparts of
GRB brighter than 1 ph cm$^{-2}$ s$^{-1}$ it is fair to conclude that
the ratio of homeless afterglows to GRB is at most a few tens, say 20. It
then follows that $\Omega_\gamma>0.05\Omega_{\rm opt}$, which combined
with our limit to $\Omega_{\rm opt}$ yields $\Omega_\gamma>0.02$. The
true rate of events that give rise to GRB is therefore at most 600 times
the observed GRB rate, and the opening angle of the ultrarelativistic,
gamma-ray emitting material is no less than $5^\circ$. Combined with the
most energetic bursts, this begins to pose a problem for the 
neutrino annihilation type of GRB energy source.

Obviously, the above calculation is only sketchy and should be taken as
an order of magnitude estimate at present. However, with the current
knowledge of afterglows a detailed calculation of the sensitivity of
the high-redshift supernova searches to GRB afterglows is feasible,
and a precise limit can be set by such a study.

\section{ Conclusions and Prospects}

Simple blast wave models seem able to accommodate the present data
on afterglows. However we can at present only infer the energy per solid 
angle; as yet the constraints on the angle-integrated $\gamma$-ray energy 
are not strong.
We must also remain aware of other possibilities. For instance, we may
be wrong in supposing that the central object becomes dormant after the
gamma-ray burst itself. It could be that the accretion-induced collapse of
a white dwarf, or (for some equations of state) the merger of two neutron 
stars, could give rise to a rapidly-spinning, temporarily rotationally 
stabilized pulsar. The afterglow could then, at least in part, be due to
a pulsar's continuing power output. It could also be that mergers of
unequal mass neutron stars, or neutron stars with other compact companions,
lead to the delayed formation of a black hole. Such events might also lead to 
repeating episodes of accretion and orbit separation, or to the eventual 
explosion of a neutron star which has dropped below the critical mass,
all of which would provide a longer time scale, episodic energy output.

We need to be open minded, yet also not too sanguine, about the
possibility of there being more subclasses of classical GRB than just
short ones and long ones. For instance, GRB with no high energy pulses
(NHE) appear to have a different (but still isotropic) spatial
distribution than those with high energy (HE) pulses (Pendleton
et~al.\ 1996). Some caution is needed in interpreting this, since
selection effects could lead to a bias against detecting HE emission in
dim bursts (Norris, 1998).  Then, there is the apparent coincidence of GRB
980425 with the SN Ib/Ic 1998bw (Galama et~al.\ 1998). A simple but
radical interpretation (Wang \& Wheeler 1998) is that all GRB may be
associated with SNe Ib/Ic and differences arise only from different
viewing angles relative to a very narrow jet. The difficulties with this
are that it would require extreme collimations by factors
$10^{-3}-10^{-4}$, and that the statistical association of any subgroup of
GRB with SNe Ib/Ic (or any other class of objects, for that matter) is so
far not significant (Kippen et~al.\ 1998). If however the GRB
980425/1998bw association is real, as argued by Woosley, Eastman \&
Schmidt (1998), Iwamoto et~al.\ (1998) and Bloom et~al.\ (1998), then we
may be in the presence of a new subclass of GRB with lower energy
$E_\gamma \sim 10^{48} (\Omj /4\pi )$ erg, which is only rarely observable
even though its comoving volume density could be substantial. In this,
more likely interpretation, the great majority of the observed GRB would
have the energies $E_\gamma \sim 10^{54}(\Omj/4\pi)$ ergs as inferred from
high redshift observations.

Much progress has been made in understanding how gamma-rays can arise in
fireballs produced by brief events depositing a large amount of energy in a 
small volume, and in deriving the generic properties of the long wavelength 
afterglows that follow from this (Rees 1998).
There still remain a number of mysteries, 
especially concerning the identity 
of their progenitors, the nature of the triggering mechanism, the transport of
the energy and the time scales involved. Nevertheless,
even if we do not yet understand the intrinsic gamma-ray burst central engine,
they may be the most powerful beacons for probing the high redshift ($z > 5$)
universe. Even if their total energy is reduced by beaming to a ``modest"
$\sim 10^{52}-10^{52.5}$ ergs in photons, they are the most extreme
phenomena that we know about in high energy astrophysics.  The modeling
of the burst itself --- the trigger, the formation of the ultrarelativistic
outflow, and the radiation processes --- is a formidable challenge to
theorists and to computational techniques. It is, also, a formidable
challenge for observers, in their quest for detecting minute details in 
extremely faint and distant sources. And if the class of models that we
have advocated here turns out to be irrelevant, the explanation of gamma-ray 
bursts will surely turn out to be even more remarkable and fascinating. 

\acknowledgements{This research has been supported by NASA NAG5-2857 and
the Royal Society. We thank H.-K. Lee for his comments and careful
reading of the manuscript.}

\end{document}